\documentclass[aps,twocolumn,superscriptaddress,longbibliography,numerical]{revtex4-1}

\usepackage{amsmath}
\usepackage{epsfig}
\usepackage{graphicx}
\usepackage{bm}
\usepackage{amssymb}
\usepackage{slashed}
\usepackage{hyperref}
\hypersetup{
     colorlinks   = true,
     citecolor    = blue,
     urlcolor = blue,
     linkcolor = blue
}

\usepackage{float} % [H] command in figures
\usepackage{caption} 
\usepackage{subcaption} % captions for subfigures

\usepackage{xcolor}
\usepackage{soul}

\usepackage{bm}
\renewcommand{\vec}[1]{\bm{#1}}

\newcommand{\integral}[2]{\int _{#1}^{#2}} % 1D integral
\newcommand{\diff}[1]{\text{d}#1} % 1D infinitesimal
\newcommand{\diffD}[2]{\text{d}^{#2}#1} % higher dimensional infinitesimal

\newcommand{\grad}[1]{\nabla #1}
\newcommand{\laplace}[1]{\nabla^2 #1}
\newcommand{\biharmonic}[1]{\nabla^4 #1}

\newcommand{\F}{\mathcal{F}}
\renewcommand{\Im}{\text{Im}}
\renewcommand{\Re}{\text{Re}}
\newcommand{\cc}{\text{c.c.}}
\renewcommand{\c}{\text{c}}
\newcommand{\Fermi}{\text{F}}
\newcommand{\FF}{\text{FF}}
\newcommand{\LO}{\text{LO}}
\newcommand{\U}{\text{U}}
\newcommand{\bulk}{\text{bulk}}
\newcommand{\boundary}{\text{surface}}

\newcommand{\e}{\text{e}} % Euler's constant

\begin{document}
% ==============================================================================

\title{Surface Pair-Density-Wave Superconducting and Superfluid States}

 \author{Mats Barkman}
 \affiliation{Department of Physics, Royal Institute of Technology, SE-106 91 Stockholm, Sweden}

 \author{Albert Samoilenka}
 \affiliation{Department of Physics, Royal Institute of Technology, SE-106 91 Stockholm, Sweden}

 \author{Egor Babaev}
 \affiliation{Department of Physics, Royal Institute of Technology, SE-106 91 Stockholm, Sweden}

% ==============================================================================
%\date{\today}
%
%
%
\begin{abstract}
Fulde, Ferrell, Larkin, and Ovchinnikov (FFLO) predicted inhomogeneous superconducting and superfluid  ground states, spontaneously breaking translation symmetries.  In this Letter, we demonstrate that the transition from the FFLO to the normal state as a function of temperature or increased Fermi surface splitting is not a direct one. Instead the system has an additional phase transition to a different state where pair-density-wave superconductivity (or superfluidity) exists only on the boundaries of the system, while the bulk of the system  is normal. The surface pair-density-wave state is very robust and exists for much larger fields and temperatures than the FFLO state. 
\end{abstract}
\maketitle

In regular BCS theory, the formation of Cooper pairs binding together two electrons with opposite spin and opposite momentum results in a uniform superconducting state \cite{Bardeen1957a,Bardeen1957b}. In 1964, Fulde and Ferrell \cite{FuldeFerrell1964} and Larkin and Ovchinnikov \cite{LarkinOvchinnikov1964} (FFLO) independently predicted that under certain conditions there should appear an inhomogeneous state in the presence of a strong magnetic field, where Zeeman splitting of the Fermi surfaces leads to the formation of Cooper pairs with nonzero total momentum. Similar inhomogeneous states can form and are of great interest in cold-atom gases \cite{Zwierlein_2006,Radzihovsky2011,Kinnunen2018,Samokhvalov2010} and in color superconducting states of quarks that are  expected to form in cores of neutron stars \cite{alford2001crystalline}.
Various predictions indicate that the FFLO state may host many rich physical phenomena including topological defects and phase transitions associated with these defects \cite{Agterberg2008,Berg2009,Agterberg2014}.
Other interesting studies include the orbital third critical magnetic field \cite{SamokhinSurface} as well as states in samples with nontrivial geometries \cite{Zyuzin2008,Zyuzin2009} and multiple competing inhomogeneous states in two-dimensional systems \cite{BarkmanMats2018IoOP}.
For a more detailed review of the FFLO state, see Refs. 
\cite{Buzdin2005,Matsuda2007,Buzdin2012}.

The anticipated interesting properties made this state highly sought after, yet
there is still no universally accepted experimental proof.
 The orbital effect is significantly stronger than the paramagnetic effect in most superconductors, hindering observations of the FFLO state. More specifically the upper critical orbital field must be significantly larger than the Chandrasekhar-Clogston limit \cite{Chandrasekhar1962,Clogston1962} for a FFLO regime to exist. Materials where possible FFLO states were discussed are heavy fermions superconductors \cite{Bianchi2003}, layered organic superconductors  \cite{Bulaevskii2003} such as $\lambda$-(BETS)$_2$FeCl$_4$ \cite{Houzet2002,Uji2006} and $\beta$'' salt \cite{Uji2018,Sugiura2019}, and iron-based superconductors \cite{Ironbased}.  Among the direct experimental probes to identify this state, it has been suggested to study the Josephson effect \cite{Yang1999} and  Andreev bound states \cite{Mayaffre2014}.
 % have been proposed to be formed in the FFLO state, which could be detected using nuclear magnetic resonance. 

 In this Letter, we report that on the phase diagrams of superconductors featuring the FFLO state should rather generically   appear another state that has a form of  surface pair-density-wave superconductivity. We find that as the Zeeman splitting field or temperature is increased, superconductivity 
 disappears only in the bulk of the system but a sample should transition into a state 
 with a superconducting surface.

The Ginzburg-Landau description of superconductors in the presence of Zeeman splitting was derived from microscopic theory in Ref. \cite{Buzdin1997}. 
The free energy functional reads $F[\psi] = \integral{\Omega}{} \F \diffD{x}{d}$ where the free energy density $\F$ is
\begin{equation} \label{eq: Free energy functional, original}
\begin{aligned}
\F =   & \alpha |\psi|^2 + \beta |\grad{\psi}|^2 + \gamma |\psi|^4 + \delta |\laplace{\psi}|^2 + \\
& \mu |\psi|^2 |\grad{\psi}|^2 + \frac{\mu}{8} \big( (\psi^*)^2(\grad{\psi})^2 + \cc \big) + \nu |\psi|^6 ,
\end{aligned}
\end{equation}
where $\psi$ is a complex field referred to as the superconducting order parameter and $\cc$ denotes complex conjugation. The coefficients $\alpha, \gamma$, and $\nu$ depend on the applied Zeeman splitting field $H$ and temperature $T$ accordingly 
\begin{align} 
\alpha & = -\pi N(0) \big(  K_1(H,T) - K_1(H_0(T),T)\big) \nonumber \\
 & \approx N(0) \frac{H-H_0(T)}{2\pi T} \Im \left[  \Psi^{(1)}\left( \frac{1}{2}- i \frac{H_0(T)}{2 \pi T}\right)\right], \label{eq: alpha coefficient} \\
\gamma  & \approx \frac{\pi N(0) K_3 (H_0(T),T)}{4}, \label{eq: gamma coefficient} \\
\nu & \approx -\frac{\pi N(0)K_5(H_0(T),T)}{8}, \label{eq: nu coefficient}
\end{align}
where $N(0)$ is the electron density of states at the Fermi surface and we have defined the functions
\begin{equation} \label{eq: K-functions}
K_n (H,T)= \frac{2T}{(2 \pi T)^n}  \frac{(-1)^n}{(n-1)!} \Re \left[  \Psi ^{(n-1)}(z)\right],
\end{equation}
where $z = 1/2- i H/2 \pi T$ and $ \Psi ^{(n)}$ is the polygamma function of order $n$. The function $H_0(T)$ indicates where $\alpha$ changes sign and is defined implicitly by the equation
\begin{equation} \label{eq: Field H_0(T)}
\ln \left( \frac{T_\c}{T}\right) = \Re \left[ \Psi^{(0)} \left(  \frac{1}{2}- i \frac{H_0(T)}{2 \pi T}\right) - \Psi^{(0)}  \left( \frac{1}{2} \right)\right],
\end{equation}
where $T_\c$ is the critical temperature above which the normal state is entered. The remaining coefficients are given as $\beta = \hat{\beta} v_\Fermi ^2 \gamma$, $\delta = \hat{\delta} v_\Fermi^4 \nu$, and $\mu = \hat{\mu} v_\Fermi^2 \nu$, where $v_\Fermi$ is the Fermi velocity and $\hat{\beta}, \hat{\delta},\hat{\mu}$ are positive constants that depend on the dimensionality $d$. In one dimension we have $\hat{\beta}=1$, $\hat{\delta}=1/2$, and $\hat{\mu}=4$ and in two dimensions we have $\hat{\beta}=1/2$, $\hat{\delta}=3/16$, and $\hat{\mu}=2$. In the parameter regime in which $\beta$ is negative, inhomogeneous order parameters may be energetically favorable. For a derivation of the Ginzburg-Landau expansion in cold atoms, see Ref. \cite{Radzihovsky2011}.

Typically considered structures of the order parameter are the so-called Fulde-Ferrell (FF) state $\psi_\FF = |\psi_\FF| \e ^{i qx}$ and the Larkin-Ovchinnikov (LO) state $\psi_\LO = |\psi_\LO| \cos qx$. For an infinite system, assuming that the order parameter vanishes close to the tricritical point, the average free energy density of these states can be minimized analytically by neglecting higher order terms, resulting in the conclusion that the LO state is energetically preferred over the FF state. The second-order phase transition into the normal state occurs at $\alpha = \alpha^\bulk_\c = \beta^2/4\delta$ . In general, the optimal order parameter structure is found by solving the equation of motion  associated with the free energy functional (called Ginzburg-Landau  equations in this context). This was done analytically in the one-dimensional case for an infinite sized superconductor, resulting in elliptic sine solutions \cite{Buzdin1997}. The sinusoidal oscillations are recovered by approaching the transition into the normal state. We solve the equation in a superconductor with boundaries. We consider the case of the real order parameter. The equation of motion can be derived through variational principles. By mapping $\psi \mapsto \psi + v$ in the free energy functional, where $v$ is some small arbitrary perturbation, we find to linear order in $v$ using Eq. \ref{eq: Free energy functional, original}
\begin{equation}
F[\psi+v] = F[\psi] + \delta F_\bulk + \delta F_\boundary %+ \mathcal{O}(v^2),
\end{equation}
where 
\begin{equation} \label{eq: dF bulk}
\begin{aligned}
\delta F_\bulk = 2 \integral{\Omega}{}   \Big(  \alpha \psi - \beta \laplace{\psi} + 2 \gamma \psi^3 + \delta \biharmonic{\psi} + \\
 \frac{5\mu}{4}\big( \psi (\grad{\psi})^2 - \psi^2 \laplace{\psi} \big) + 3 \nu \psi^5 \Big)v \diffD{x}{d}
\end{aligned}
\end{equation}
and
\begin{equation} \label{eq: dF boundary}
\begin{aligned}
\delta F_\boundary = 2 \integral{\partial \Omega}{} & \Bigg\lbrace \Big(  \Big[\beta + \frac{5\mu}{4}\psi^2 \Big] \grad{\psi} - \delta\nabla^3 \psi\Big)v  \\
& + \Big(  \delta  \laplace{\psi}\Big) \grad{v}\Bigg\rbrace \cdot \vec{n} \diff{S},
\end{aligned}
\end{equation}
where $\vec{n}$ is the normal vector to the boundary $\partial \Omega$. By setting $\delta F_\bulk = 0$ we find the equation of motion and by setting $\delta F_\boundary = 0$ we find the two boundary conditions
\begin{align}
\Big(\Big[\beta + \frac{5\mu}{4}\psi^2 \Big] \grad{\psi} - \delta\nabla^3 \psi \Big) \cdot \vec{n} = 0, \label{eq: BC1}\\
\delta \laplace{\psi} = 0. \label{eq: BC2}
\end{align}

It is convenient to rescale the theory in the regime where $\beta$ is negative by defining the dimensionless quantities $\tilde{\psi} = \psi / |\psi_\U|$, $\tilde{\alpha} = \alpha/\alpha_\U $, $\tilde{x} = q_0 x $, where $|\psi_\U|^2 = -\gamma/2\nu$, $\alpha_\U = \gamma^2/4\nu $, and $q_0^2 = -\beta/2\delta$. The free energy can thus be written $F[\psi] = \alpha_\U |\psi_\U|^2/q_0^d  \tilde{F}[\tilde{\psi}]$, where $\tilde{F}[\tilde{\psi}] = \integral{\Omega}{} \tilde{\F} \diffD{\tilde{x}}{d}$, in which the rescaled free energy density is identical to Eq. \ref{eq: Free energy functional, original}, where the coefficients have been replaced accordingly $\alpha \mapsto \tilde{\alpha}, \beta \mapsto \tilde{\beta}$, and so on, where $\tilde{\gamma}=-2\tilde{\nu} = -2$, $\tilde{\beta} = -2\tilde{\delta} = -2 \hat{\beta}^2 / \hat{\delta}$, and $\tilde{\mu} = \hat{\beta} \hat{\mu}/\hat{\delta}$. Consequently, there is only one free parameter $\tilde{\alpha}$ in the rescaled theory to vary, which parametrizes changes in both temperature and Zeeman splitting field.

\begin{figure*}
\center
\begin{subfigure}[t]{0.45\textwidth}
\center
\includegraphics[height=9cm]{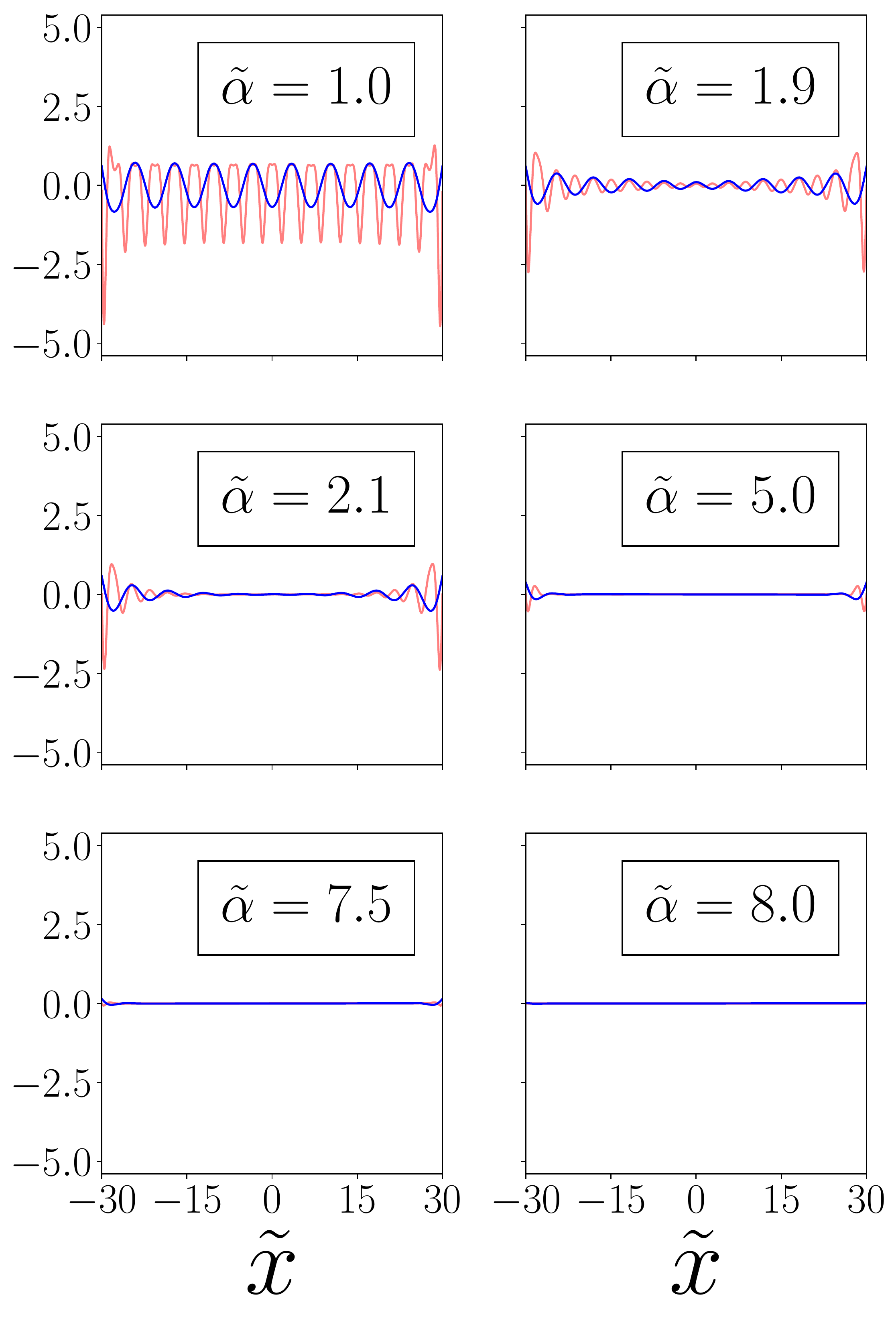}
\end{subfigure}
\begin{subfigure}[t]{0.5\textwidth}
\center
\includegraphics[height=9cm]{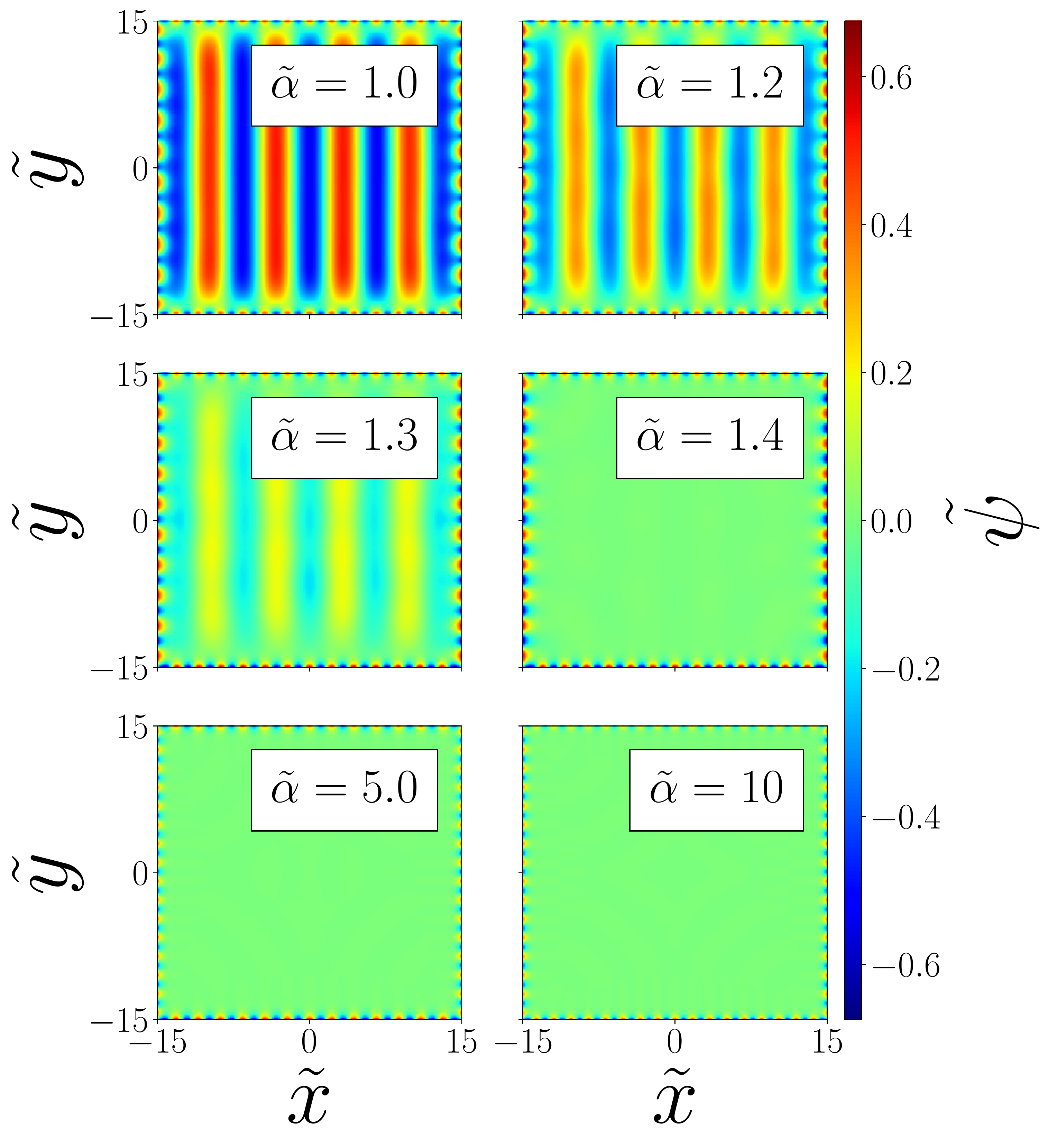}
\end{subfigure}
\caption{Numerically calculated order parameters $\tilde{\psi}$ in one- and two-dimensional system domains for various temperatures or, equivalently, various Zeeman splitting fields, parametrized by $\tilde{\alpha}$. In the one-dimensional domain, the blue curve is the order parameter and the red curve is the energy density. One can clearly observe a nontrivial oscillatory pattern of the energy density in the bulk of the system. The images show the sequence for going from superconducting to normal state by increasing the parameter $\tilde{\alpha}$. In the first panel, the system is in the FFLO state, and in the last panel the system is close to the normal state. For intermediate $\tilde{\alpha}$, the order parameter first vanishes in the bulk of the system, and the system enters the surface pair-density-wave state that persists for a much wider range of parameters than the FFLO state. In calculations of the two-dimensional system, the term $0.5|\tilde{\nabla} \tilde{\psi}|^4$ was retained in the Ginzburg-Landau expansion, for details see Ref. \cite{Samoilenka}.} \label{fig: Order parameters}
\end{figure*}

Having derived boundary conditions, we will now numerically minimize the free energy for a superconductor in both one- and two-dimensional domains, while varying $\tilde{\alpha}$. The associated free energy is calculated in order to locate phase transitions.  Two different numerical approaches were used. The solutions in Fig. \ref{fig: Order parameters} were obtained using a finite difference scheme and nonlinear conjugate gradient method, parallelized on a graphical processing unit. These results were also supported by calculations using the finite element method within  FreeFem++ framework \cite{FreeFem}.

% The obtained order parameters for various $\tilde{\alpha}$ are shown in Fig. \ref{fig: Order parameters}.

We find that the free energy remains negative for $\tilde{\alpha}$ larger than the critical value $\tilde{\alpha}^\bulk_\c$, where in one dimension $\tilde{\alpha}^\bulk_\c=2$ and in two dimensions $\tilde{\alpha}^\bulk_\c=4/3$. The origin of it is the formation of a distinct  surface pair-density-wave (PDW) superconducting state,
which has a superconducting gap on the boundaries of the system but not in its bulk. The obtained order parameter structure has the form of a damped oscillation with an amplitude that vanishes in the bulk but remains nonzero close to the boundaries. The boundary states are found in both one- and two-dimensional systems. The results have been verified by altering the system size and it was found that,
for a sufficiently large system, the surface  state is independent of system size. 

The origin of the appearance of the surface PDW state is the following: besides the inhomogeneous order parameter, the bulk FFLO state has inhomogeneous 
energy density. The numerical solutions for  one- and two-dimensional cases are plotted in 
 Fig. \ref{fig: Order parameters}. As the system approaches the phase transition from the bulk FFLO to the bulk normal
 state, the energy gain from the areas with negative energy density becomes balanced by the areas with positive energy density.
 However, if the system has a boundary, a solution can  be found where the boundary cuts off a segment of inhomogeneous solutions with positive energy and has a decaying oscillatory configuration of the order parameter extending to a certain length scale in the bulk. That is, a decaying solution near the boundary 
 starting with a negative energy segment should be stable even when the system does not support the FFLO state in the bulk. 
Indeed, the numerical solutions clearly show that the boundaries are characterized by negative energy density as seen in Fig. \ref{fig: Order parameters}, resulting in the stability of the surface PDW state for large $\tilde{\alpha}$.
The generality of the argument implies that these surface superconducting states of nontopological origin should be present for all states where  free energy density is oscillating in space. This includes generalizations of FFLO states to systems with unconventional pairing \cite{BarkmanZyuzin}.

The existence of a surface PDW state in a semi-infinite system $\Omega = [0,\infty )$ 
can be proven analytically.
To that end, we use an variational ansatz of the form
\begin{equation} \label{eq: Parametrization analytical guess}
\psi (x) = \Delta  \e ^{-kx} \cos (qx + \phi),
\end{equation}
where the parameters $\Delta$, $k$, $q$ and $\phi$ should be found such that the free energy is minimized,
subject to the  boundary conditions.
The surface PDW to normal transition occurs when the energy is minimal with $\Delta=0$.
When the transition into the normal state is of second order, it is sufficient to consider terms proportional to $|\Delta|^2$ in the free energy. In addition, the boundary condition in Eq.~\ref{eq: BC1} takes the simpler form $(\beta \grad{\psi} - \delta \nabla^3 \psi) \cdot \vec{n}= 0$ close to the transition into the normal state. Carrying out the minimization analytically, we find that the free energy associated with the surface PDW state remains negative until $\alpha = \alpha^\boundary_\c = 4\alpha_\c ^\bulk$. 
The numerical calculation shows that the phase transition indeed occurs at this value of $\alpha$, which implies that the variational ansatz in Eq. \ref{eq: Parametrization analytical guess} captures very well the solution for the surface PDW state in one dimension.

%Since this is an ansatz-based analytical calculation, it should
%underestimate the critical temperature. The numerical results indicate that the transition to the surface superconductivity occurs at only about a 20 \%  larger value of  $\alpha$, which implies  that the variational ansatz in Eq. \ref{eq: Parametrization analytical guess}   captures fairly well the solution for surface PDW state.

%

\begin{figure}[H]
\center
\includegraphics[width=0.4\textwidth]{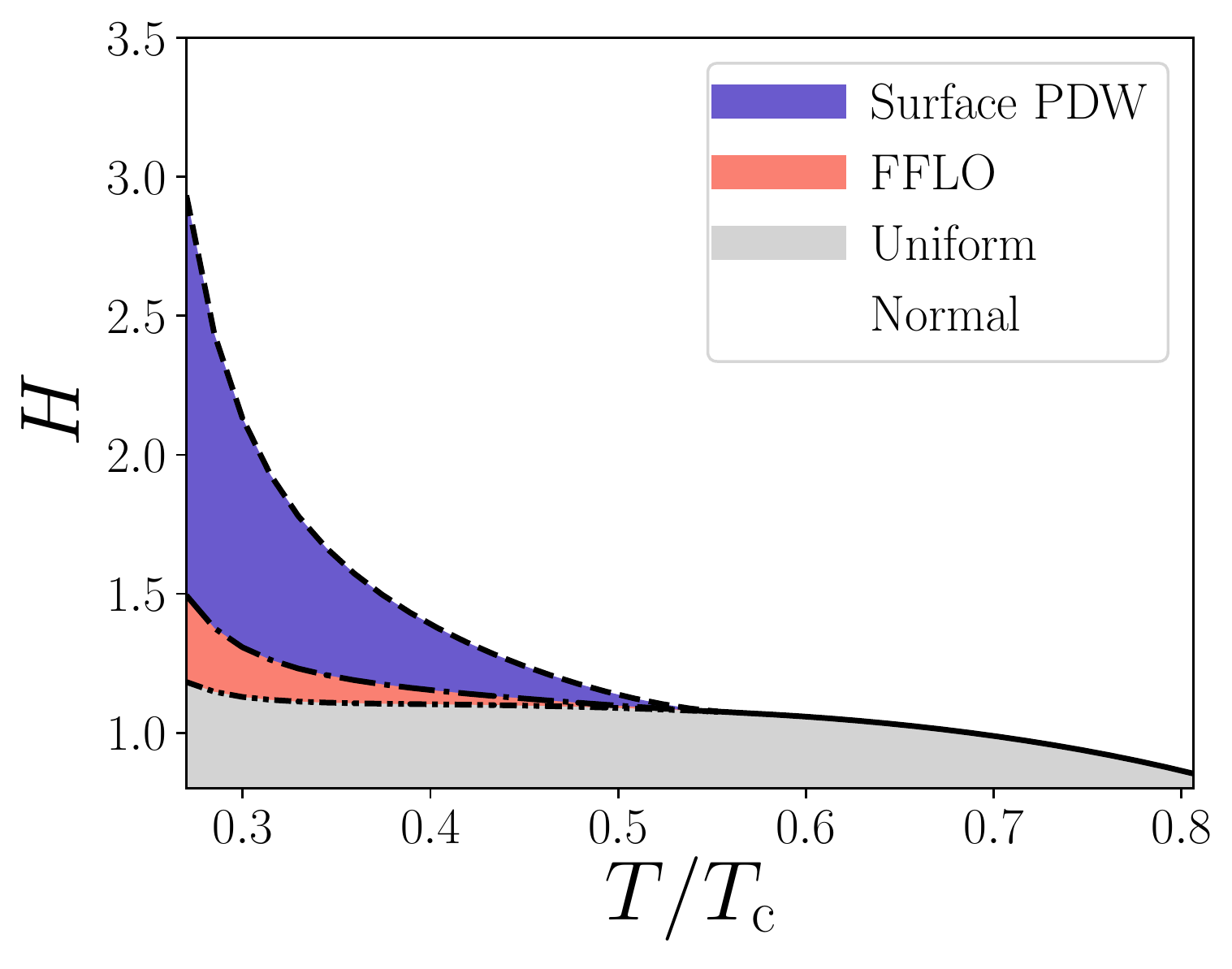}
\caption{Phase diagram for a superconductor with boundaries in the presence of Zeeman splitting of Fermi surfaces caused by the in plane external field $H$. The role of the external field here is to create imbalance of electronic populations with different spins. In the context of cold atoms, this corresponds to fermionic population imbalance. At low $H$ the superconductor is in the uniform superconducting state. At lower temperatures and high $H$, the inhomogeneous regime is entered. There exists a narrow regime in which the FFLO state exists in the bulk of the superconductor. One finds a large surface PDW state regime where the system has superconducting boundaries but normal bulk. The figure shows calculations based on a one dimensional
model.} \label{fig: Phase diagram}
\end{figure}

We can draw the phase diagram with respect to $H$ and $T$ for the one-dimensional system, as shown in Fig. \ref{fig: Phase diagram}. The inhomogeneous regime is split into two parts: the bulk FFLO state and the surface PDW  state. The  regime of surface PDW  state  on the phase diagram is significantly larger than the bulk FFLO regime. 

Note that the only role of the external magnetic field is to create an electronic population imbalance.
For example, in two dimensions the situation corresponds to an in plane field and there is no external field perpendicular to the plane.  
In the context of cold atoms, it corresponds to the
fermionic population imbalance in the absence of rotation. Therefore, the physical origin and the structure of the solution is very different from that of the third upper critical magnetic field $H_{\c 3}$ \cite{SAINTJAMES1963} that was recently studied in FFLO systems \cite{SamokhinSurface}. The results also have implications for the problem of FFLO states in mesoscopic systems, 
where the literature  focuses on commensurateness effects \cite{Zyuzin2008,Zyuzin2009,BdGSmall}.
%If we gradually decrease a system size  in our solutions, towards the mesoscopic limit
 % the areas of the surface PDW superconductivity will start overlapping.
%If we consider temperatures where bulk state does not exist, then taking our solution and gradually decreasing the size of systems, the surface areas will start to overlap.
Consider the parameter regime where the surface PDW state exists. By gradually decreasing the system size, the areas with the PDW state that live on opposite surfaces will eventually start to overlap with each other.
Thus, the above results imply that, for a range of mesoscopic system sizes of the order of $1/k$ in Eq. \ref{eq: Parametrization analytical guess}, the most robust solutions will have a form very different from a periodically oscillating function, which warrants further investigation.

In conclusion, we have studied 
superconducting or superfluid systems supporting the FFLO state.
We found that, at elevated temperatures or for strong splitting of Fermi surfaces,
such systems support a state
where the surface of the sample is a PDW superconductor or  
 \noindent
superfluid while the bulk 
of the system is normal. Correspondingly, when the temperature is increased, the system has two phase transitions:
Firstly, superconductivity disappears in the bulk of the system, while surfaces remain superconducting. Secondly, the system transitions into the fully normal state only at a higher temperature.
In the considered regime, we found that the surface PDW state is more robust than the FFLO state and extends to much larger values of Fermi surface splitting and temperatures. 

The effect can be used to experimentally prove the elusive
FFLO state as follows: The main specific heat feature should be detectable well below the superconducting phase transition. This is because the contribution to the specific heat should be  dominated by the bulk, where the gap disappears earlier than on the surface.
% In this case the main specific heat feature should occur well below superconducting phase transition: becasue it should be dominated by the
%  disappearance of the   gap in the bulk.
The transport measurements should at the same time indicate that system retains superconductivity due to the surface PDW state.
Because of superconductivity existing only in a thin layer
in the surface PDW state, another expected experimental signature is a concomitant increase of magnetic field
penetration lengths for fields perpendicular to the superconducting layer. Very sensitive specific heat measurements should see two features associated with the bulk and surface phase transitions and yield the phase diagram shown in Fig. \ref{fig: Phase diagram}.
In cold atoms \cite{Radzihovsky2011}, the surface PDW state
can be directly observed   for experiments realizing sharp potential walls \cite{box1,box2}.
Finally, the results have implications for the models of color superconductivity in the neutron stars cores, at the interface between nuclear and quark matter, which
is widely believed to be of the FFLO type \cite{alford2001crystalline}. 
%The results imply that analogue of the surface color superconducting PDW state may form on the boundary between color superconducting quark matter and dense nuclear matter.

The work was supported by the Swedish Research
Council Grants No. 642-2013-7837 and No.  VR2016-06122 and Goran  Gustafsson  Foundation  for  Research  in  Natural  Sciences  and  Medicine.  
 The computations were performed on resources provided by the
Swedish National Infrastructure for Computing (SNIC) at the National Supercomputer Center
 in Linkoping, Sweden.

\bibliography{references.bib}

\end{document}